\documentstyle[manuscript,prl,aps,epsfig]{revtex}

\catcode`@=11
\def\citer{\@ifnextchar [{\@tempswatrue\@citexr}{\@tempswafalse\@citexr[]}}
 
\def\@citexr[#1]#2{\if@filesw\immediate\write\@auxout{\string\citation{#2}}\fi
  \def\@citea{}\@cite{\@for\@citeb:=#2\do
    {\@citea\def\@citea{--\penalty\@m}\@ifundefined
       {b@\@citeb}{{\bf ?}\@warning
       {Citation `\@citeb' on page \thepage \space undefined}}%
\hbox{\csname b@\@citeb\endcsname}}}{#1}}
\catcode`@=12

\def\beq{\begin{equation}}
\def\eeq{\end{equation}}

\def\beqn{\begin{eqnarray}}
\def\eeqn{\end{eqnarray}}
\relax
\def\ba{\begin{array}}
\def\ea{\end{array}}

\begin{document}                                                              
\thispagestyle{empty}
\null
\hfill KA--TP--29--1996\\[-.2em]
\null 
\hfill MPI--PhT/96--129\\[-.2em]
\null
\hfill PM/96--36\\[-.2em]
\null
\hfill hep-ph/9612363\\
\vskip .8cm
\begin{center}
{\Large \bf Supersymmetric Contributions to Electroweak\\[.4em] 
Precision Observables: QCD Corrections%
\footnote{Work supported by the Deutsche Forschungsgemeinschaft.}}
\vskip 2.5em
{\large
{\sc A.~Djouadi$^{1,2}$, P. Gambino$^{3}$, S. Heinemeyer$^2$, W.
Hollik$^2$, \\[.2em]
C.~J\"unger$^2$ and G. Weiglein$^2$}}\\[1ex]
{\normalsize \it 
$^1$ Physique Math\'ematique et Th\'eorique, 
Universit\'e de Montpellier II,\\
F--34095 Montpellier Cedex 5, France \\
$^2$ Institut f\"ur Theoretische Physik, Universit\"at Karlsruhe,
D--76128 Karlsruhe, Germany \\
$^3$ Max--Planck Institut f\"ur Physik, Werner Heisenberg Institut,
D--80805 Munich, Germany}
\vskip 2em
\end{center} \par
\vskip 1.2cm
\vfil
{\bf Abstract} \par
We calculate the two--loop QCD correction to the scalar quark
contributions to the electroweak gauge boson self--energies 
at zero momentum--transfer in the supersymmetric extension of the
Standard Model. We then derive the ${\cal O}(\alpha_s)$ 
correction to the contribution of the scalar top and bottom quark
loops to the $\rho$ parameter, which is the most sizable supersymmetric 
contribution to the electroweak mixing angle and the $W$--boson mass. 
The two--loop corrections modify the one--loop contribution 
by up to 30\%; the gluino decouples for large masses.
Contrary to the SM case where the QCD corrections 
are negative and screen the one--loop value, the corresponding 
corrections in the supersymmetric case are in general positive,  
increasing the sensitivity in the search for scalar quarks through their 
virtual effects in high--precision electroweak observables. 
\par
\vskip 1cm
\null
\setcounter{page}{0}




Supersymmetric theories (SUSY)~\cite{R1} are the best motivated
extensions of the Standard Model (SM) of the electroweak and strong
interactions. They provide an elegant way to break the electroweak
symmetry and to stabilize the huge hierarchy between the GUT and the
Fermi scales, and allow for a consistent unification of the gauge
coupling constants as well as a natural solution of the Dark Matter
problem; for recent reviews see Ref.~\cite{R2}. 

Supersymmetry predicts the
existence of scalar partners to each SM fermion, and spin--1/2 partners
to the gauge and Higgs bosons. So far, the direct search of SUSY
particles at colliders has not been successful, and under some
assumptions one can only set lower bounds of ${\cal O}(100)$ GeV on their
masses~\cite{R3}. The search can be extended to slightly larger values at 
LEP2 and the upgraded Tevatron; higher energy hadron or $e^+e^-$ colliders 
will be required to sweep the entire range of the SUSY particle masses 
up to the TeV scale. 

An alternative way to probe SUSY is to search for
the virtual effects of the additional particles. Indeed, now that the
top--quark mass --- the measured value of which being in remarkable
agreement with the predicted one --- is known~\cite{R3}, one can use the 
high--precision electroweak data to search for the quantum effects 
of the SUSY particles: sfermions, charginos/neutralinos and gluinos. 

In the Minimal Supersymmetric Standard Model (MSSM) it is well known 
that, besides the rare decay $b \to s \gamma$~\cite{R3a},
there are two possibilities for the virtual effects of SUSY 
particles to be large enough to be detected in present high--precision 
experiments. The first possibility is that charginos and scalar top 
quarks are light enough to affect the decay width of the $Z$ boson into 
$b$--quarks~\cite{R4}; however, for masses beyond the LEP2 or Tevatron 
reach, these effects become too small to be observable~\cite{R5}. 

The second possibility is the contribution of the scalar top and bottom 
quark loops to the electroweak gauge--boson self--energies~\cite{R6}: if 
there is a large splitting between the masses of these particles, the 
contribution will grow with the mass of the heaviest scalar quark and 
can be sizable. This is similar to the SM case, where the top/bottom 
weak isodoublet generates a quantum correction that grows as $m_t^2$. 
This contribution enters the electroweak observables via the $\rho$ 
parameter~\cite{R7}, which measures the relative strength of the neutral 
to charged current processes at zero momentum--transfer. It is mainly from 
this contribution that the top--quark mass has been successfully predicted 
from the measurement of the effective electroweak mixing angle at the 
$Z$--boson resonance and the $W$--boson mass at hadron colliders~\cite{R3}, a 
triumph for the electroweak theory. 

In order to treat the SUSY loop contributions to the electroweak
observables at the same level of accuracy as the standard contribution,
higher order corrections should be incorporated. In particular the QCD 
corrections, which because of the large value of the strong coupling 
constant can be rather important, must be known. It is the purpose of 
this report to provide the two--loop QCD corrections to the scalar quark 
contributions to the electroweak precision observables. As a first step,
we will consider here only the contributions to the $\rho$ parameter; 
more detailed results will be given elsewhere~\cite{R8}. 

\smallskip 

The $\rho$ parameter, in terms of the transverse parts of 
the $W$-- and $Z$--boson self--energies at zero momentum--transfer, 
is given by 
\beq
\rho = \frac{1}{1-\Delta \rho} \ ; \ \Delta \rho = 
\frac{\Pi_{ZZ}(0)}{M_Z^2} - \frac{\Pi_{WW}(0)}{M_W^2}  \ .
\eeq
In the SM, the contribution of a fermion isodoublet $(u,d)$
to $\Delta \rho$ reads at one--loop order
\beq
\Delta \rho_0^{\rm SM} = \frac{N_c G_F}{8 \sqrt{2} \pi^2} F_0 
\left( m_u^2, m_d^2 \right) \ , 
\eeq
with the color factor $N_c$ and the function $F_0$ given by
\beq
F_0(x,y)= x+y - \frac{2xy} {x-y} \log \frac{x}{y} \ . 
\eeq
The function $F_0$ vanishes if the $u$-- and $d$--type quarks are 
degenerate in mass: $F_0(m_q^2, m_q^2)=0$; in the limit of large quark 
mass splitting it becomes proportional to the heavy quark mass 
squared: $F_0(m_q^2,0)=m_q^2$. Therefore, in the SM the only relevant
contribution is due to the top/bottom weak isodoublet. Because 
$m_t \gg m_b$, one obtains $\Delta \rho ^{\rm SM}_0 = 3 G_Fm_t^2/(8 
\sqrt{2} \pi^2)$, a large contribution which allowed for the prediction
of $m_t$. However, in order that the predicted value agrees with the 
experimental one, QCD corrections have to be included. These two--loop 
corrections have been calculated ten years ago, leading to a 
result~\cite{R7a}:
$\Delta \rho ^{\rm SM}_1 = - \Delta \rho_0^{\rm SM} \cdot 
\frac{2}{3} \frac{\alpha_s}{\pi} (1+\pi^2/3 )$. For the value $\alpha_s 
\simeq 0.12$, the QCD correction \cite{R7b} decreases the one--loop result 
by approximately $10\%$ and shifts $m_t$ upwards by an amount of $\sim 10$~GeV. 

In SUSY theories, the scalar partners of each SM quark will induce 
additional contributions. The current eigenstates, 
$\tilde{q}_L$ and $\tilde{q}_R$, mix to give the mass eigenstates. 
The mixing angle is proportional to the quark mass and therefore is 
important only in the case of the third generation scalar quarks~\cite{R9}. 
In particular, due to the large value of $m_t$, the mixing angle 
$\theta_{\tilde{t}}$ between $\tilde{t}_L$ and $\tilde{t}_R$ can be 
very large and lead to a scalar top quark  $\tilde{t}_1$ much 
lighter than the $t$--quark and all the scalar partners of the light 
quarks~\cite{R9}. The mixing in the $\tilde{b}$--quark sector can be
sizable only in a small area of the SUSY parameter space. 

The contribution of a scalar quark doublet $(\tilde{u}, \tilde{d})$
to the transverse parts of the $W/Z$--boson self--energies at zero 
momentum--transfer [Fig.~1] can be written as~\cite{R6}
\begin{eqnarray}
\Pi_{WW} (0)
& = & - \frac{3 G_F M_W^2}{8 \sqrt{2} \pi^2} \, \sum_{i,j=1,2} 
\left( a^{\tilde{u}}_i a^{\tilde{d}}_j \right)^2 \, F_0 \left( 
m_{\tilde{u}_i}^2, m_{\tilde{d}_j}^2 \right), \nonumber \\
\Pi_{ZZ}(0)
& = &- \frac{3G_F M_Z^2}{8 \sqrt{2} \pi^2} \, \frac{1}{2} 
\sum_{\tilde{q}= \tilde{u}, \tilde{d} \atop i,j =1,2}
(a_i^{\tilde{q}} a_j^{\tilde{q}} )^2 F_0 \left( 
m_{\tilde{q}_i}^2, m_{\tilde{q}_j}^2 \right) , 
\end{eqnarray}
where the factors $a_{i}^{\tilde{q}}$ are given in terms of the scalar
quark mixing angle $\theta_{\tilde{q}}$ as $a_{1}^{\tilde{q}}= \cos \theta_{
\tilde{q}}$ and $a_{2}^{\tilde{q}} = \sin \theta_{\tilde{q}}$. 


As can be seen from $F_0$ in eq.~(3), the contribution of a scalar 
quark doublet vanishes if all masses are  degenerate. This means that 
in most SUSY scenarios, where the scalar partners of the light quarks 
are almost mass degenerate, only the third generation will contribute.
Neglecting the mixing in the $\tilde{b}$ sector, $\Delta \rho$ is given 
at one--loop order by the simple expression
\beqn
\Delta \rho ^{\rm SUSY}_0 &=& \frac{3 G_F}{8 \sqrt{2} \pi^2} \left[ -
\sin^2 \theta_{\tilde{t}} \cos^2 \theta_{\tilde{t}} 
F_0\left( m_{\tilde{t}_1}^2,  m_{\tilde{t}_2}^2 \right)
\right. \nonumber \\ 
&& \left. + \cos^2 \theta_{\tilde{t}} F_0\left( m_{\tilde{t}_1}^2,  
m_{\tilde{b}_L}^2 \right) + \sin^2 \theta_{\tilde{t}} F_0\left( 
m_{\tilde{t}_2}^2,  m_{\tilde{b}_L}^2 \right) \right]. 
\eeqn
In a large area of the parameter space, the scalar top mixing angle is 
either very small $\theta_t \sim 0$ or maximal, $\theta_t \sim -\pi/4$. 
The contribution $\Delta \rho_0^{\rm SUSY}$ is shown in Fig.~2
as a function of the common scalar mass $m_{\tilde{q}}=m_{\tilde{t}_{L,R}}
=m_{\tilde{b}_{L}}$ for these two scenarios. The contribution can be at the
level of a few per mille and therefore within the range of the experimental
observability. Relaxing the assumption of a common scalar quark mass, 
the corrections can become even larger~\cite{R6}. 


\smallskip 

At ${\cal O}(\alpha \alpha_s)$, the two--loop Feynman diagrams contributing
to the $\rho$ parameter in SUSY [Fig.~3] 
consist of two sets which, at vanishing 
external momentum and after the inclusion of the counterterms, are separately 
ultraviolet finite and gauge-invariant. The first one has 
diagrams involving only 
gluon exchange, Fig.~3a; in this case 
the calculation is similar to the SM, although technically more complicated 
due to the larger number of diagrams and the presence of $\tilde{q}$ mixing. 
The diagrams involving the quartic scalar--quark interaction in Fig.~3a
will either contribute only to the longitudinal component of the 
self--energies or can be absorbed into the $\tilde{q}$ mass and mixing angle 
renormalization as will be discussed later. The second set consists of 
diagrams involving scalar quarks, gluinos as well as quarks, 
Fig.~3b; in this case the calculation becomes very complicated due to 
the even larger number of diagrams and to the presence of up to 5 particles 
with different masses in the loops. 


We have calculated the two--loop contribution of a complete quark/squark 
generation to the vacuum polarization functions of the electroweak gauge 
bosons at zero momentum--transfer, taking into account  general mixing 
between scalar quarks and allowing for all particles to have 
different masses. In the following, we summarize the main features of the 
calculation~\cite{R14}.

Our results have been derived by two independent calculations using
different methods. In one method, the unrenormalized self--energies 
together with the mass and mixing angle counterterms were calculated with 
the help of the program {\it ProcessDiagram}~\cite{XX}, while in the other 
case the packages {\it FeynArts} \cite{R11} [in 
which the relevant part of the MSSM has been implemented] and
{\it TwoCalc} \cite{R12} were used 
to generate and evaluate the full set of Feynman diagrams and 
counterterms. The two independent calculations allowed for thorough checks 
of the final results.

The two--loop Feynman diagrams of Fig.~3 have to be supplemented by
the corresponding counterterm insertions into the one--loop diagrams. 
By virtue of the Ward identity, the vertex and wave--function
renormalization constants cancel each other. The mass renormalization
has been performed in the on--shell scheme, where the mass is
defined as the pole of the propagator. The mixing angle renormalization 
is performed in such a way that all transitions from $\tilde{q}_i 
\leftrightarrow \tilde{q}_j$ which do not depend on the loop--momenta 
in the two--loop diagrams are canceled; this renormalization condition 
is equivalent to the one proposed in Ref.~\cite{R13} for scalar 
quark decays. With this choice of the mass and mixing angle renormalization, 
the pure scalar quark diagrams in Fig.~3a that contribute to the 
transverse parts of the gauge--boson self--energies are canceled. 

In order to discuss our results, let us first concentrate on the 
contribution of the gluonic corrections, 
Fig.~3a, and the corresponding counterterms.
At the two--loop level, the results for the electroweak gauge 
boson self--energies at zero momentum--transfer have very simple 
analytical expressions. In the case of an isodoublet $(\tilde{u}, 
\tilde{d})$ where general mixing is allowed, the structure is similar 
to eq.~(4) with the $a_i^{\tilde{q}}$ as given previously:
\beqn
\Pi_{WW} (0)
& = & - \frac{G_F M_W^2 \alpha_s}{4 \sqrt{2} \pi^3} \, \sum_{i,j=1,2} 
\left( a^{\tilde{u}}_i a^{\tilde{d}}_j \right)^2 \, F_1 \left( 
m_{\tilde{u}_i}^2, m_{\tilde{d}_j}^2 \right), \nonumber \\
\Pi_{ZZ}(0)
& = &- \frac{G_F M_Z^2 \alpha_s}{8 \sqrt{2} \pi^3}   
\sum_{\tilde{q}= \tilde{u},\tilde{d} \atop i,j=1,2} (a_i^{\tilde{q}} 
a_j^{\tilde{q}} )^2 \, F_1 \left( m_{\tilde{q}_i}^2, m_{\tilde{q}_j}^2 
\right) .
\end{eqnarray}
The two--loop function $F_1(x,y)$ is given in terms of dilogarithms by
\beqn
F_{1}(x,y) &=& x+y- 2\frac{xy}{x-y} \log \frac{x}{y} \left[2+
\frac{x}{y} \log \frac{x}{y} \right] \nonumber \\
&& {} +\frac{(x+y)x^2}{(x-y)^2}\log^2 \frac{x}{y} 
-2(x-y) {\rm Li}_2 \left(1-\frac{x}{y} \right) . 
\eeqn
This function is symmetric in the interchange of $x$ and $y$.
As in the case of the one--loop function $F_0$, it vanishes for 
degenerate masses, $F_1(x,x)=0$, while in the case of large 
mass splitting it increases with the heavy scalar quark mass 
squared: $F_1 (x,0) = x( 1 +\pi^2/3)$.  

{}From the previous expressions, the contribution of the $(\tilde{t}, 
\tilde{b})$ doublet to the $\rho$ parameter, including the two--loop 
gluon exchange and pure scalar quark diagrams are obtained 
straightforwardly. In the case where the $\tilde{b}$ mixing is neglected, 
the SUSY two--loop contribution is given by an expression similar to 
eq.~(5):
\beqn
\Delta \rho ^{\rm SUSY}_1 &=& \frac{G_F \alpha_s}{4 \sqrt{2} \pi^3} \left[ 
- \sin^2\theta_{\tilde{t}} \cos^2\theta_{\tilde{t}}  
F_1\left( m_{\tilde{t}_1}^2,  m_{\tilde{t}_2}^2 \right) \right. \nonumber \\ 
&&\left. + \cos^2 \theta_{\tilde{t}} F_1 \left( m_{\tilde{t}_1}^2,  
m_{\tilde{b}_L}^2 \right)
+\sin^2 \theta_{\tilde{t}}  F_1 \left( 
m_{\tilde{t}_2}^2,  m_{\tilde{b}_L}^2 \right) \right]. 
\eeqn
The two--loop gluonic SUSY contribution to $\Delta \rho$ is shown in Fig.~4 
as a function of the common scalar mass $m_{\tilde{q}}$, for the two 
scenarios discussed previously: $\theta_{\tilde{t}} = 0$ and 
$\theta_{\tilde{t}} \simeq -\pi/4$. As can be seen, the two--loop contribution 
is of the order of 10 to 15\% of the one--loop result. Contrary to the SM 
case [and to many QCD corrections to electroweak processes in the SM, see
Ref.~\cite{GK} for a review] where the two--loop correction screens 
the one--loop contribution, $\Delta \rho_1^{\rm SUSY}$ has the same sign 
as $\Delta \rho_0^{\rm SUSY}$. For instance, in the case of degenerate 
$\tilde{t}$ quarks with masses $m_{\tilde{t}} \gg m_{\tilde{b}}$, the 
result is the same as the QCD correction to the $(t,b)$ contribution 
in the SM, but with opposite sign. The gluonic correction to the 
contribution of scalar quarks to the $\rho$ parameter will therefore 
enhance the sensitivity in the search of the virtual effects of scalar
quarks in high--precision electroweak measurements. 


The analytical expressions of the contribution of the two--loop diagrams 
with gluino exchange, Fig.~3b, to the electroweak gauge boson self--energies
are very complicated even at zero momentum--transfer. Besides the
fact that the scalar quark mixing leads to a large number of 
contributing diagrams, this is mainly due to the presence of up 
to five particles with different masses in the loops. The lengthy expressions 
will be given elsewhere~\cite{R8}.  It turned out that in general 
the gluino exchange diagrams give smaller contributions compared to gluon 
exchange.  Only for gluino and scalar quark masses close to the 
experimental lower bounds they compete with the gluon exchange 
contributions. In this case, the gluon and gluino contributions add
up to $\sim 30\%$ of the one--loop value for maximal mixing [Fig.~5].
For larger values of $m_{\tilde{g}}$, the contribution
decreases rapidly since the gluinos decouple for high masses. 


Finally, let us note that for the diagrams in Fig.~3a analytical
expressions for arbitrary momentum--transfer can be obtained as will 
be discussed in Ref.~\cite{R8}. With the present computational knowledge 
of two--loop radiative corrections, analytical exact results for the 
diagrams involving gluino exchange, Fig.~3b, cannot be obtained for 
arbitrary $q^2$; either approximations like heavy mass 
expansions or numerical methods have to be applied. 

In summary, we have calculated the two--loop ${\cal O}(\alpha_s)$ correction 
to the scalar quark contributions to the weak gauge boson self--energies at 
zero momentum--transfer in SUSY theories, and derived the QCD 
correction to the $\rho$ parameter. The gluonic corrections are of  
${\cal O}(10\%)$: they are positive and increase the sensitivity in the search 
for scalar quarks through their virtual effects in high--precision electroweak 
observables. The gluino contributions are in general smaller except for
relatively light gluinos and scalar quarks; the contribution vanishes for 
large gluino masses. The phenomenological implications of our results will 
be discussed in a forthcoming paper.


\vspace{1 cm}

\begin{figure}[htb]
\begin{center}
\mbox{
\psfig{figure=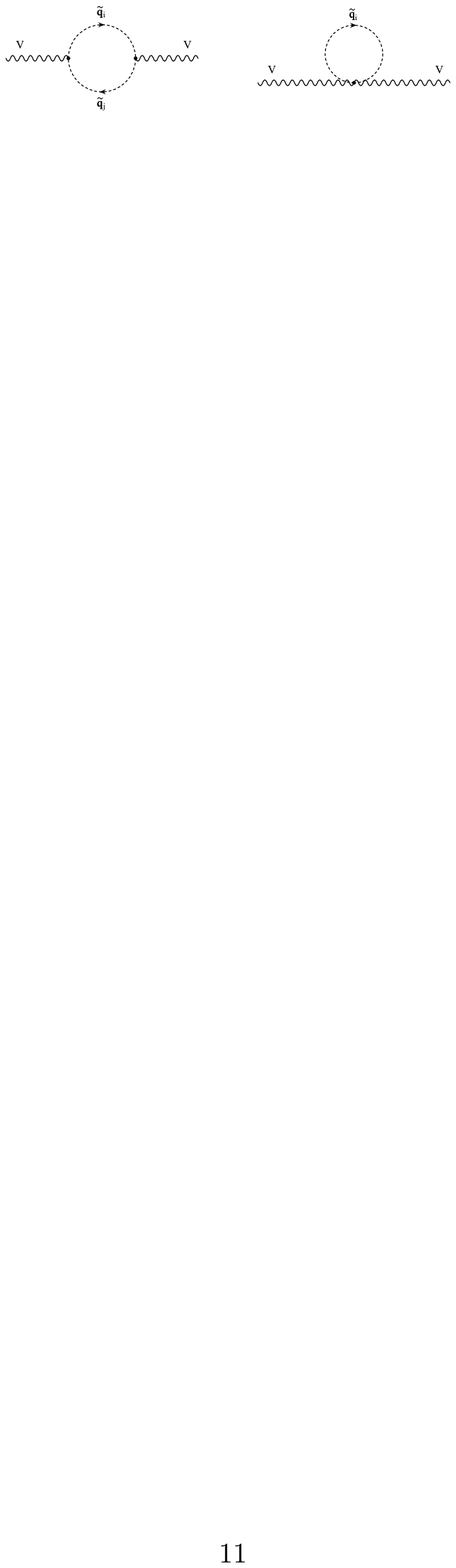,width=12cm,bbllx=210pt,bblly=680pt,bburx=397pt,bbury=720pt}}
\end{center}
\caption[]{Feynman diagrams for the contribution of scalar quark loops 
to the gauge boson self--energies at one--loop.}
\end{figure}

\vspace{2 cm}

\begin{figure}[htb]
\begin{center}
\mbox{
\psfig{figure=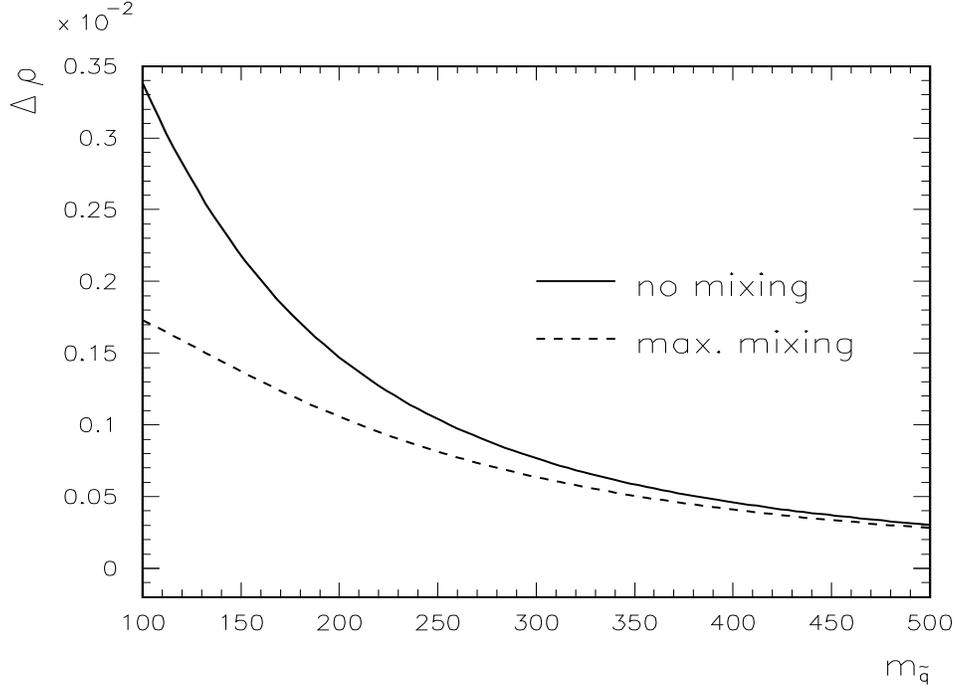,width=12cm,height=8.5cm,bbllx=140pt,bblly=285pt,bburx=450pt,bbury=535pt}}
\end{center}
\caption[]{One--loop contribution of the $(\tilde{t}, \tilde{b})$ 
doublet to $\Delta \rho$ as a function of the common mass $m_{\tilde{q}}$, 
for $\theta_{\tilde{t}} =0$ and $\theta_{\tilde{t}} \sim-\pi/4$
[with $\tan \beta=1.6$ and $m_{\rm LR}=0$ and 200 GeV, respectively, 
where $m_{\rm LR}$ is the off--diagonal term in the $\tilde{t}$ mass matrix].}
\end{figure}

\begin{figure}[htb]
\begin{center}
\mbox{
\psfig{figure=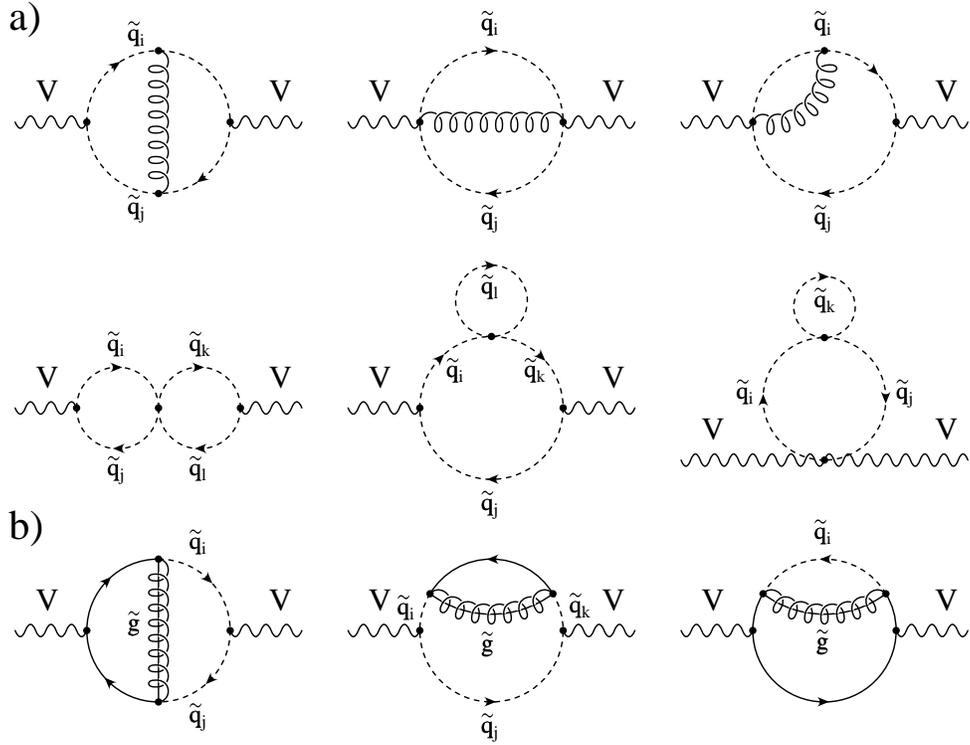,width=12cm,bbllx=165pt,bblly=495pt,bburx=450pt,bbury=725pt}}
\end{center}
\caption[]{Typical Feynman diagrams for the contribution of scalar 
quarks and gluinos to the $W/Z$--boson self--energies at the 
two--loop level.}
\end{figure}

\clearpage

\begin{figure}[htb]
\begin{center}
\mbox{
\psfig{figure=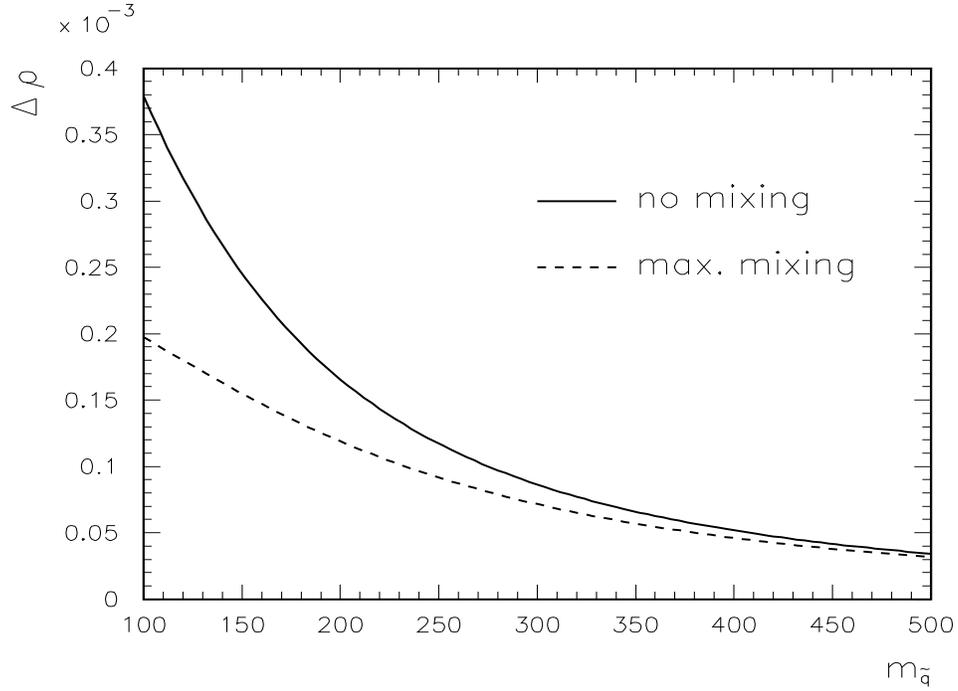,width=12cm,height=8.5cm,bbllx=140pt,bblly=285pt,bburx=450pt,bbury=535pt}}
\end{center}
\caption[]{Gluon exchange contribution to the $\rho$ parameter at two--loop 
as a function of $m_{\tilde{q}}$ for the scenarios of Fig.~2.}
\end{figure} 

\begin{figure}[htb]
\begin{center}
\mbox{
\psfig{figure=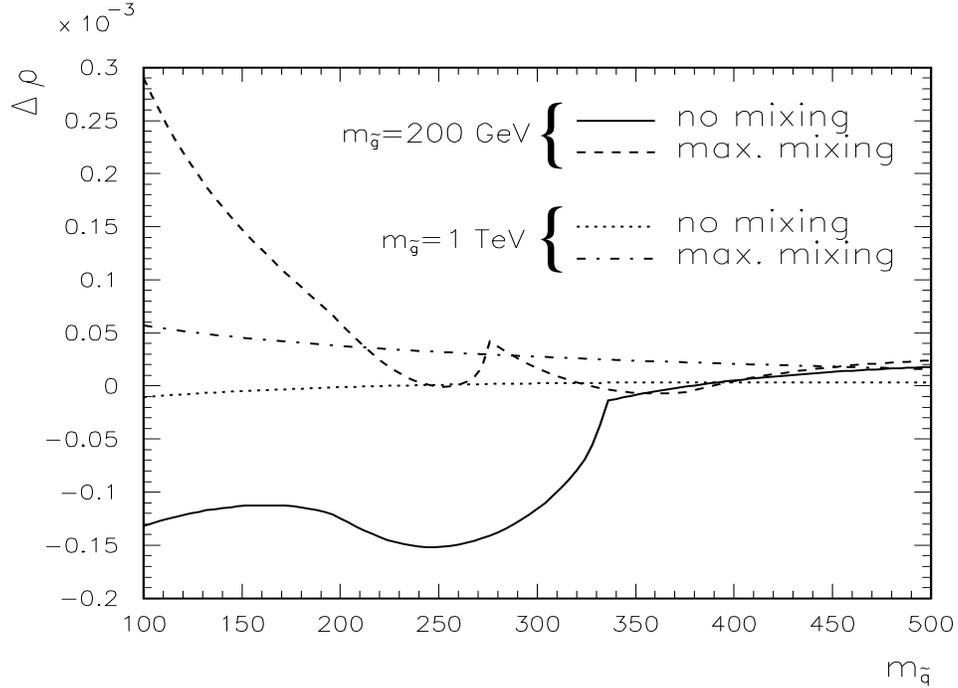,width=12cm,height=8.5cm,bbllx=140pt,bblly=285pt,bburx=450pt,bbury=535pt}}
\end{center}
\caption[]{Contribution of the gluino exchange diagrams to $\Delta \rho_1^
{\rm SUSY}$ for two values of $m_{\tilde{g}}$ in the scenarios of Fig.~2.}
\end{figure}

\end{document}